\documentclass[twocolumn,usenatbib]{aastex62}

\usepackage{graphicx}	% Including figure files
\usepackage{amsmath}	% Advanced maths commands
\usepackage{amssymb}	% Extra maths symbols

\received{\today}
\revised{\today}
\accepted{\today}
\submitjournal{ApJ}

\shorttitle{Nonthermal particle acceleration in turbulence}
\shortauthors{Zhdankin et al.}
\begin{document}

\newcommand{\red}{\textcolor{red}}
\newcommand{\blue}{\textcolor{blue}}
\newcommand{\green}{\textcolor{green}}

%\title{Nonthermal particle acceleration in large-scale relativistic plasma turbulence}
\title{System-size convergence of nonthermal particle acceleration in relativistic plasma turbulence}

\correspondingauthor{Vladimir Zhdankin}
\email{zhdankin@jila.colorado.edu}

\author{Vladimir Zhdankin}
\affiliation{JILA, NIST and University of Colorado, 440 UCB, Boulder, Colorado 80309, USA}

\author{Dmitri A. Uzdensky}
\affiliation{Center for Integrated Plasma Studies, Department of Physics, 390 UCB, University of Colorado, Boulder, CO 80309, USA}

\author{Gregory R. Werner}
\affiliation{Center for Integrated Plasma Studies, Department of Physics, 390 UCB, University of Colorado, Boulder, CO 80309, USA}

\author{Mitchell C. Begelman}
\affiliation{JILA, NIST and University of Colorado, 440 UCB, Boulder, Colorado 80309, USA}
\affiliation{Department of Astrophysical and Planetary Sciences, 391 UCB, Boulder, CO 80309, USA}

%\nocollaboration

%% Note that the \and command from previous versions of AASTeX is now
%% depreciated in this version as it is no longer necessary. AASTeX 
%% automatically takes care of all commas and "and"s between authors names.

%% AASTeX 6.2 has the new \collaboration and \nocollaboration commands to
%% provide the collaboration status of a group of authors. These commands 
%% can be used either before or after the list of corresponding authors. The
%% argument for \collaboration is the collaboration identifier. Authors are
%% encouraged to surround collaboration identifiers with ()s. The 
%% \nocollaboration command takes no argument and exists to indicate that
%% the nearby authors are not part of surrounding collaborations.

%% Mark off the abstract in the ``abstract'' environment. 
\begin{abstract}
We apply collisionless particle-in-cell simulations of relativistic pair plasmas to explore whether driven turbulence is a viable high-energy astrophysical particle accelerator. We characterize nonthermal particle distributions for varying system sizes up to $L/2\pi\rho_{e0} = 163$, where $L/2\pi$ is the driving scale and $\rho_{e0}$ is the initial characteristic Larmor radius. We show that turbulent particle acceleration produces power-law energy distributions that, when compared at a fixed number of large-scale dynamical times, slowly steepen with increasing system size. We demonstrate, however, that convergence is obtained by comparing the distributions at different times that increase with system size (approximately logarithmically). We suggest that the system-size dependence arises from the time required for particles to reach the highest accessible energies via Fermi acceleration. The converged power-law index of the energy distribution, $\alpha \approx 3.0$ for magnetization $\sigma = 3/8$, makes turbulence a possible explanation for nonthermal spectra observed in systems such as the Crab nebula.
\end{abstract}

%% Keywords should appear after the \end{abstract} command. 
%% See the online documentation for the full list of available subject
%% keywords and the rules for their use.
\keywords{acceleration of particles, magnetohydrodynamics (MHD), plasmas, pulsars: individual (Crab), relativistic processes, turbulence}

%% From the front matter, we move on to the body of the paper.
%% Sections are demarcated by \section and \subsection, respectively.
%% Observe the use of the LaTeX \label
%% command after the \subsection to give a symbolic KEY to the
%% subsection for cross-referencing in a \ref command.
%% You can use LaTeX's \ref and \label commands to keep track of
%% cross-references to sections, equations, tables, and figures.
%% That way, if you change the order of any elements, LaTeX will
%% automatically renumber them.
%%
%% We recommend that authors also use the natbib \citep
%% and \citet commands to identify citations.  The citations are
%% tied to the reference list via symbolic KEYs. The KEY corresponds
%% to the KEY in the \bibitem in the reference list below. 

\section{Introduction} \label{sec:intro}

For many decades, turbulence has been recognized as a conceivable source of nonthermal energetic particles in collisionless plasmas. Theoretical works have proposed a variety of routes toward particle acceleration, including diffusive (second-order) acceleration from turbulent fluctuations [Alfv\'{e}nic modes \citep{jokipii_1966, schlickeiser_1989, chandran_2000, cho_lazarian_2006}; compressive modes \citep{schlickeiser_miller_1998, yan_lazarian_2002, chandran_2003}; kinetic modes \citep[e.g.,][]{dermer_etal_1996, fonseca_etal_2003, petrosian_liu_2004, riquelme_etal_2017}] and secular (first-order) acceleration via intermittent structures [shocks \citep{bykov_toptygin_1982, blandford_eichler_1987}; current sheets undergoing magnetic reconnection \citep{vlahos_etal_2004, lazarian_etal_2012, isliker_etal_2017}; see also \cite{beresnyak_li_2016}]. These mechanisms of acceleration are tantalizing theoretical possibilities, but rely on various assumptions about the nature of turbulence and nonlinear plasma physics. Due the complexity and analytic intractability of the problem, the only practical way to prove the reality of turbulent particle acceleration (apart from direct experimental confirmation) is with self-consistent, large-scale numerical simulations.

Turbulent particle acceleration has important implications for space systems such as the solar corona, the solar wind, and planetary magnetospheres, as well as for high-energy astrophysical systems such as pulsar wind nebulae, X-ray binaries, supernovae remnants, jets from active galactic nuclei (including blazars), radio lobes, and gamma ray bursts. Observations of broadband radiation spectra and cosmic rays imply that nonthermal particles are a significant component of the universe. In this work, we focus on plasmas that are relativistically hot with modestly relativistic bulk velocities, as found in many high-energy astrophysical settings.

In our previous work \citep{zhdankin_etal_2017}, we applied particle-in-cell (PIC) simulations to demonstrate that driven turbulence can produce a substantial population of nonthermal particles in relativistic pair plasmas, with power-law energy distributions that become harder with increasing magnetization (ratio of magnetic enthalpy to relativistic plasma enthalpy). However, these simulations also revealed that the distributions became softer with increasing system size, and were therefore unable to probe distributions in the asymptotic large-system limit. In principle, a lack of convergence can arise from inadequate scale separation, or from the adverse role of physical effects such as scale-dependent anisotropy, intermittency, damping of relevant (e.g., compressive) modes, or the inherent inefficiency of the acceleration process at magnetohydrodynamic (MHD) scales. Since supercomputers will be unable to simulate systems with sizes comparable to real astrophysical systems in the foreseeable future, it is necessary to understand the scaling of nonthermal particle distributions with system size before applying such simulations to model astrophysical phenomena.

In the present work, we address the system-size dependence of turbulent particle acceleration. We confirm a weak system size dependence for nonthermal energy distributions when measured at a fixed number of large-scale dynamical times, for sizes extending beyond those considered in \cite{zhdankin_etal_2017}. However, more importantly, we present evidence that the distributions converge when compared at different times that increase with system size (approximately logarithmically or as a weak power law). Physically, this time dependence arises from the fact that the distributions do not fully develop until particles reach the highest accessible energies via Fermi acceleration. The converged value of the index for the power-law energy distribution ($\alpha \approx 3.0$ for magnetization $\sigma = 3/8$) confirms turbulence as an efficient, viable astrophysical particle accelerator.

\section{Simulations} \label{sec:intro}

We perform the simulations with the explicit electromagnetic PIC code {\sc Zeltron} \citep{cerutti_etal_2013} using charge-conserving current deposition \citep{esirkepov_2001}. The simulation set-up is described in detail in \cite{zhdankin_etal_2018}; here, we simply outline the main features. The domain is a periodic cubic box of size $L^3$ (consisting of $N^3$ cells) with uniform mean magnetic field $\boldsymbol{B}_0 = B_0 \hat{\boldsymbol{z}}$. We initialize electrons and positrons from a uniform Maxwell-J\"{u}ttner distribution with combined particle density $n_0$ and temperature $T_0 = \theta_0 m c^2$, where $m$ is the electron rest mass, and we choose $\theta_0 = 100$ (giving an ultra-relativistic initial mean Lorentz factor of $\gamma_0 \approx 300$). We then drive strong ($\delta B_{\rm rms} \sim B_0$) turbulence at low wavenumber modes ($k = 2\pi/L$) by applying a randomly fluctuating external current density \citep{tenbarge_etal_2014}. For these simulations, we fix the initial magnetization to $\sigma_0 \equiv B_0^2/4\pi h_0 = 3/8$, where $h_0 = 4 n_0 \theta_0 m c^2$ is the initial relativistic enthalpy density. The initial Alfv\'{e}n velocity is given by $v_{A0} \equiv c [\sigma_0/(\sigma_0 + 1)]^{1/2} \approx 0.52 c$; we perform each simulation for a duration of at least $7.5 L/v_{A0}$. As optimized by convergence studies, we set the initial Larmor radius to $\rho_{e0} \equiv \gamma_0 m c^2/ e B_0 = 1.5 \Delta x$ (where $\Delta x$ is the cell size) and choose $64$ particles per cell for the main simulations.

We perform a scan over system size $L/2\pi\rho_{e0}$ by varying the number of cells in each  simulation, taking~$N \in \{ 256, 384, 512, 768, 1024, 1536 \}$, so that~$L/2\pi\rho_{e0} \in \{ 27.2, 40.7, 54.3, 81.5, 109, 163 \}$. To check reproducibility, we reran all of the cases having $N \le 1024$ with a different random seed (for particle initialization and driving phases); for robustness, we analyze the particle distributions averaged for each simulation pair at a given size. We also perform statistical ensembles of sixteen $384^3$ cases and eight $768^3$ cases with 32 particles per cell to investigate statistical variation of the results.

\section{Results} \label{sec:results}

\begin{figure}
  \includegraphics[width=\columnwidth]{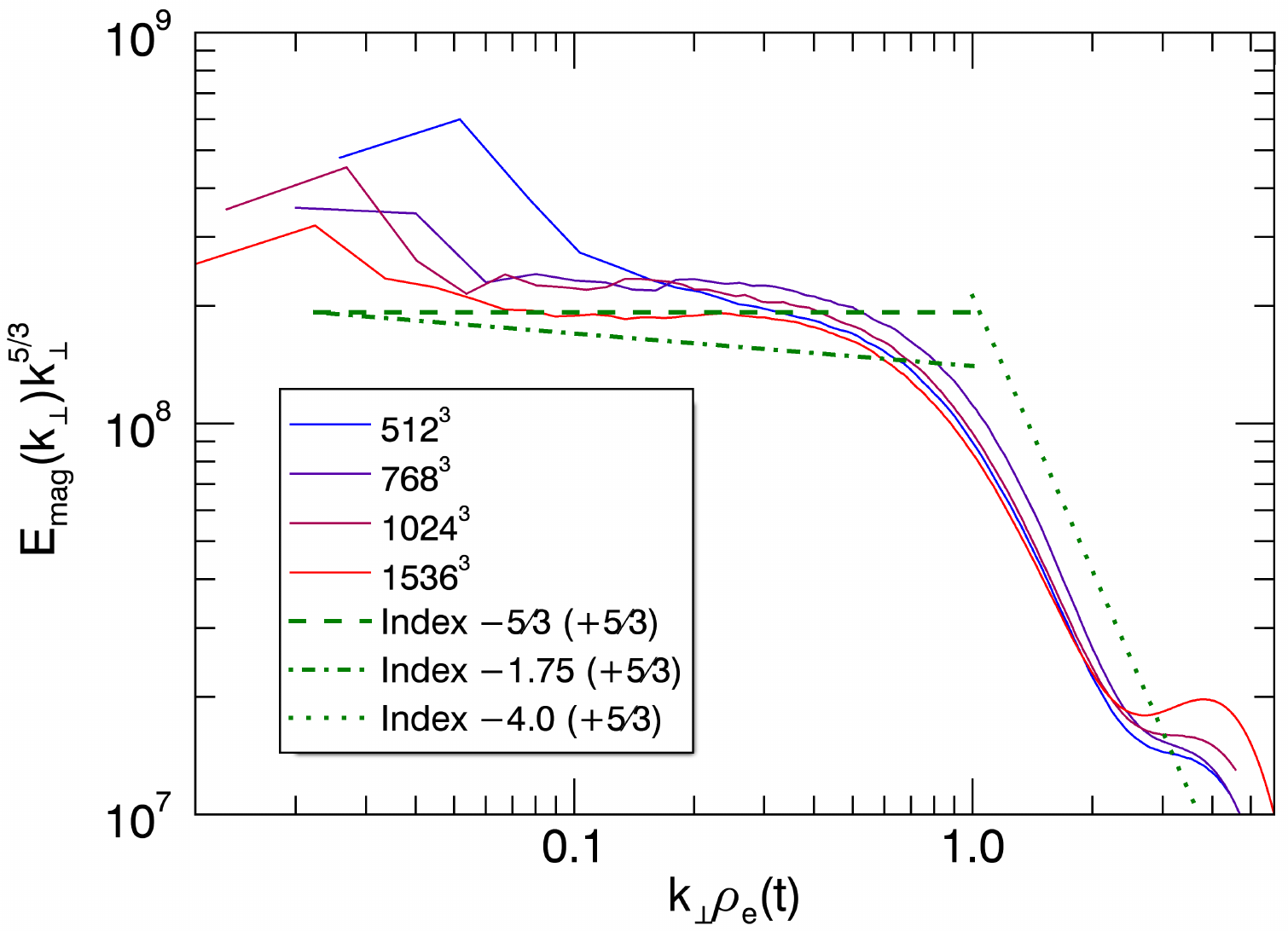}
   \centering
   \caption{\label{fig:spec_mag} Magnetic energy spectrum compensated by $k_\perp^{5/3}$ for varying system sizes. Power laws with pre-compensated indices of $-5/3$ (dashed), $-1.75$ (dash-dotted), and $-4$ (dotted) are shown for reference.}
 \end{figure}

The time evolution of the simulations proceeds as discussed in our previous papers \citep{zhdankin_etal_2017, zhdankin_etal_2018}: the external driving disrupts the initial thermal equilibrium and establishes turbulent fluctuations across a broad range of scales. The turbulence is fully developed after a few Alfv\'{e}n times, after which turbulent energy dissipation increases the internal energy at a constant rate. For reference, in Fig.~\ref{fig:spec_mag}, we show the magnetic energy spectrum compensated by $k_\perp^{5/3}$, where $k_\perp$ is the wavenumber perpendicular to $\boldsymbol{B}_0$, for simulations of varying size, averaged over 5 snapshots from $3.1 \le t v_{A0}/L \le 5.2$. Whereas the $512^3$ case has a spectrum that is steeper than $k_\perp^{-5/3}$, the larger cases ($768^3$ and above) have spectra close to $k_\perp^{-5/3}$, in agreement with classical MHD turbulence theories \citep{goldreich_sridhar_1995, thompson_blaes_1998}. The $1536^3$ case exhibits an inertial range from $k_\perp \rho_e \sim 0.06$ to $k_\perp \rho_e \sim 0.4$, where $\rho_e = \langle\gamma\rangle mc^2/eB_{\rm rms}$ is the characteristic Larmor radius based on the instantaneous mean particle Lorentz factor $\langle\gamma\rangle$.
 
 \begin{figure}
  \includegraphics[width=\columnwidth]{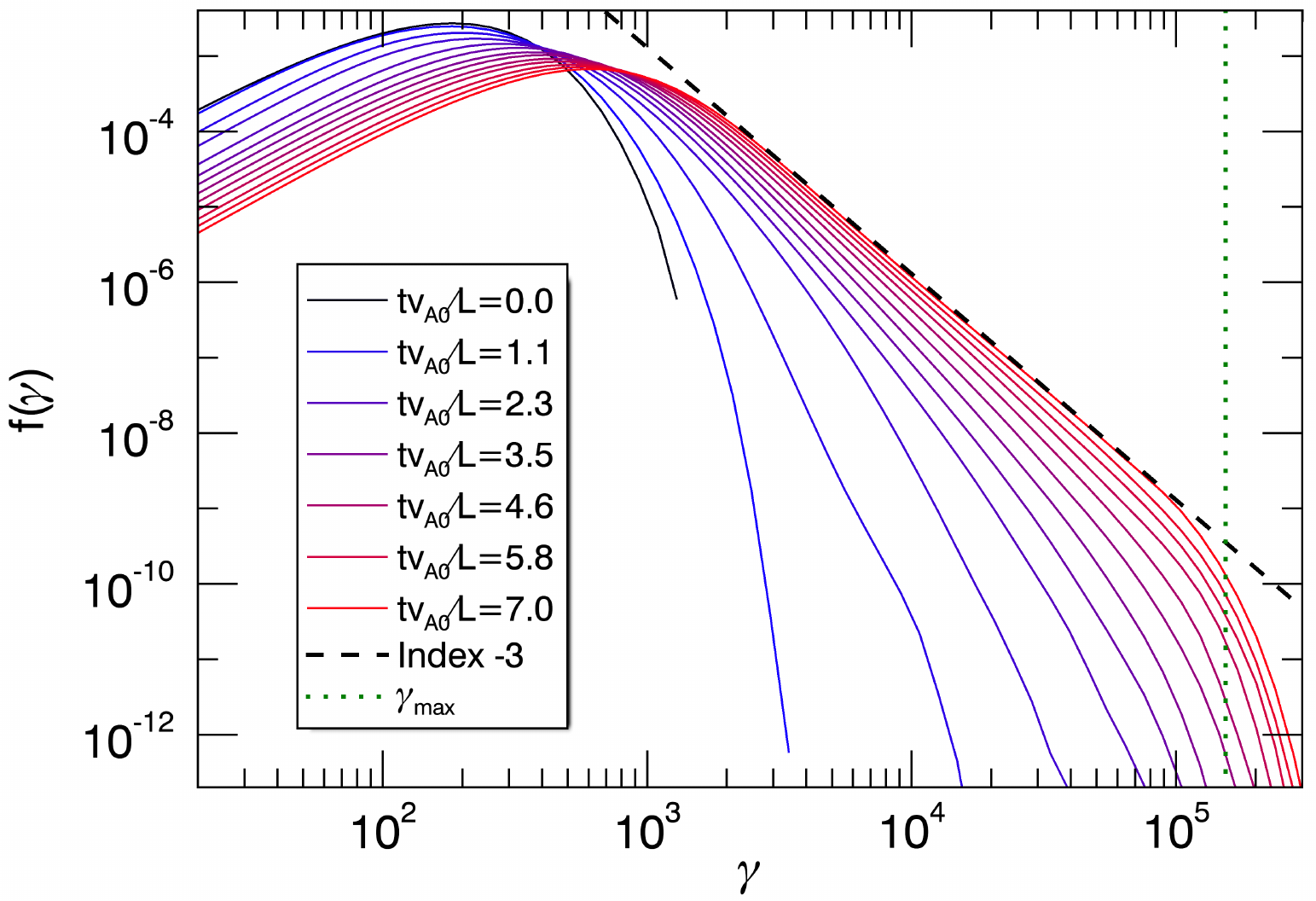}
    \includegraphics[width=\columnwidth]{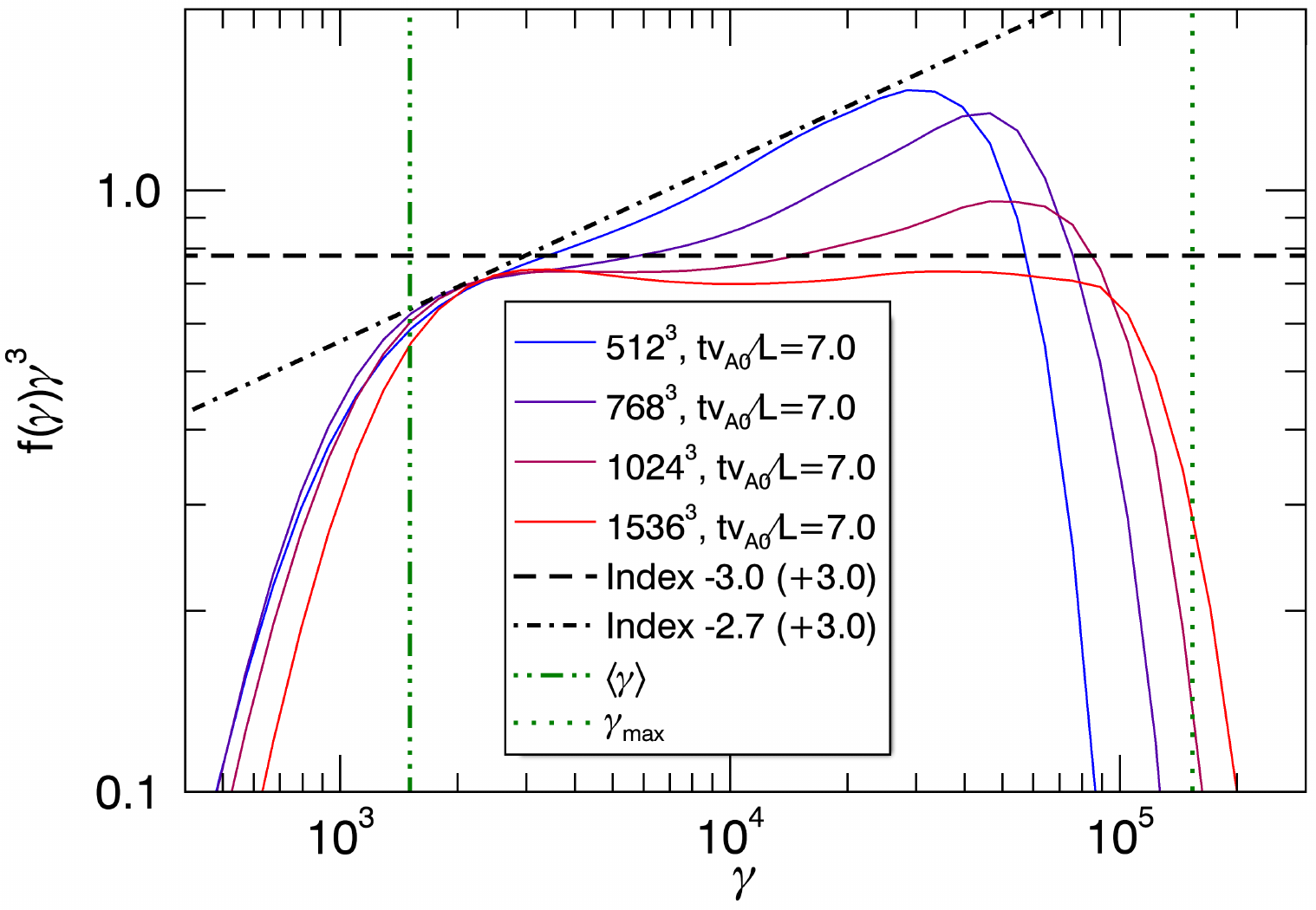}
   \includegraphics[width=\columnwidth]{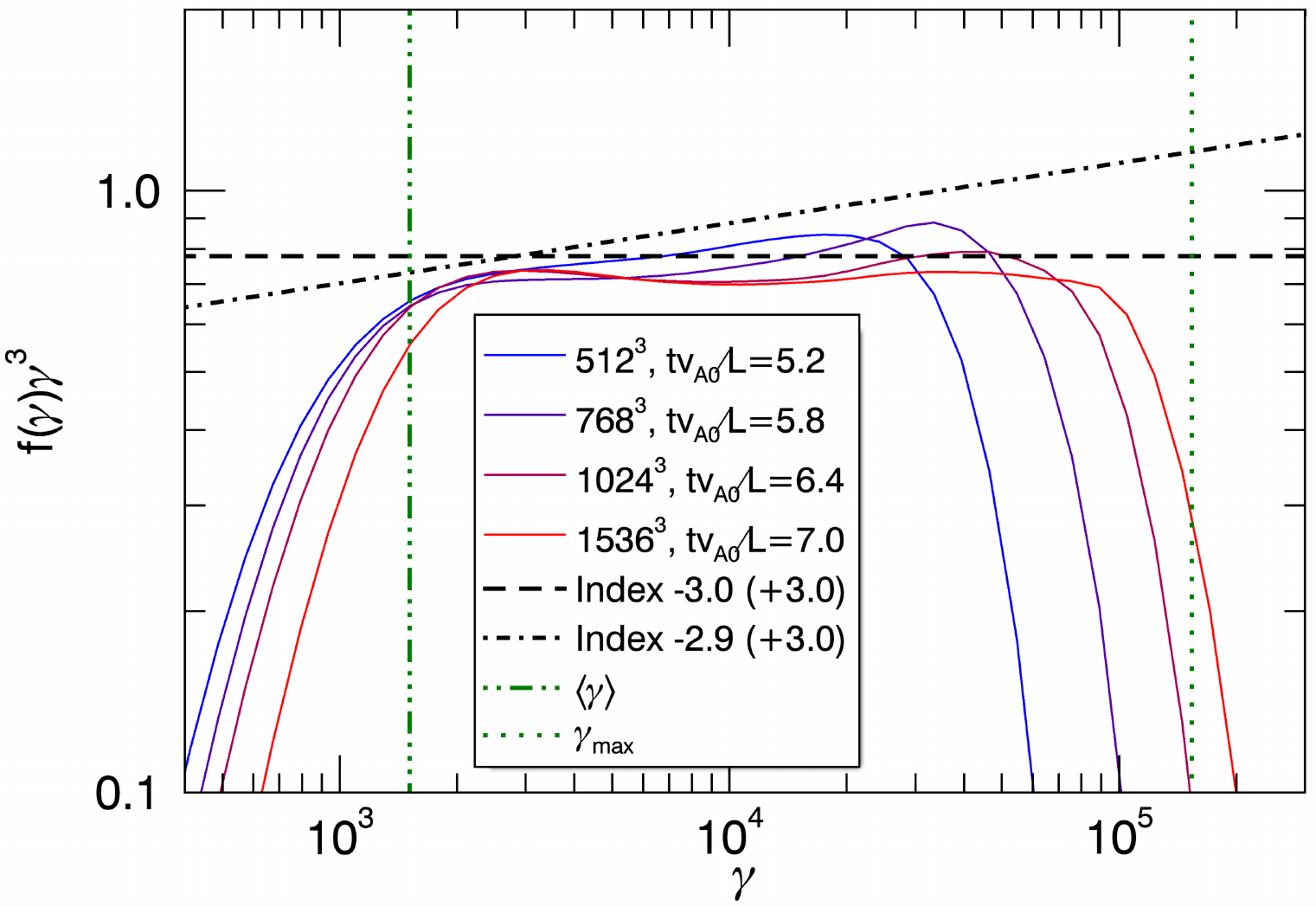}
   \centering
   \caption{\label{fig:dist} Top panel: Evolution of the particle energy distribution $f(\gamma)$ for the $1536^3$ simulation. Center panel: Compensated distribution $f(\gamma)\gamma^3$, at fixed time $tv_{A0}/L = 7.0$, for varying system sizes. Power laws with pre-compensated index $-3.0$ (black dashed) and $-2.7$ (black dash-dotted) are also shown, along with the mean energy $\langle\gamma\rangle$ (green dash-dotted) and system-size cutoff $\gamma_{\rm max}$ (green dotted) for the $1536^3$ case. Bottom panel: Similar compensated distributions at times increasing logarithmically with size. Power laws with pre-compensated index $-3.0$ (black dashed) and $-2.9$ (black dash-dotted) are shown in this case.}
 \end{figure}
 
We now turn to the particle energy distribution $f(\gamma)$, where the particle Lorentz factor $\gamma = E/mc^2$ is used interchangeably with energy $E$. The evolution of $f(\gamma)$ for the $1536^3$ simulation is shown in the top panel of Fig.~\ref{fig:dist}. As in our previous work \citep{zhdankin_etal_2017}, the distribution develops a power-law tail, $f(\gamma) \sim \gamma^{-\alpha}$, over several dynamical times, attaining an index of $\alpha \approx 3.0$ at $t v_{A0}/L \sim 7$. This power law extends from energies comparable to the instantaneous mean ($\langle\gamma\rangle \sim 1.5 \times 10^3$) up to energies limited by the system size ($\gamma_{\rm max} \equiv LeB_0/2mc^2 \sim 1.5 \times 10^5$), extending across a factor of $\sim 50$ in energy. At later times (not shown), particles accumulate at energies near $\gamma_{\rm max}$, causing a high-energy pileup in the distribution.

In the middle panel of Fig.~\ref{fig:dist}, we show the energy distributions at a fixed number of large-scale dynamical times, taken to be $t v_{A0}/L = 7.0$, for simulations of varying size ($512^3$ to $1536^3$). For clarity, we compensate the distributions by $\gamma^3$, making the $1536^3$ case horizontal. The nonthermal tail steepens with increasing system size, ranging from an estimated index of $\alpha \approx 2.7$ for the $512^3$ case to $\alpha \approx 3.0$ for the $1536^3$ case. Thus, when compared at fixed times, there is no clear evidence for convergence of $f(\gamma)$ with system size; although the scaling of $\alpha$ with size is weak ($\delta\alpha \sim 0.3$ for a factor of $3$ increase in size), it can undermine the viability of turbulent particle acceleration in astrophysical systems if it persists to larger sizes.

The interpretation of the data changes, however, when the energy distributions are compared at different times, chosen to scale with system size. In the bottom panel of Fig.~\ref{fig:dist}, we show distributions at $tv_{A0}/L \in \{5.2, 5.8, 6.4, 7.0\}$ for the simulations with $\{512^3, 768^3, 1024^3, 1536^3\}$ cells (which is an approximately logarithmic increase of time with size). When compared at these times, the $768^3$, $1024^3$, and $1536^3$ simulations all exhibit converged distributions with index near $\alpha \approx 3.0$, to within $\pm 0.1$ accuracy. Notably, these times approximately coincide with the initial formation of the pileup at $\gamma_{\rm max}$. This leads to our main proposal, that \it turbulent particle acceleration produces a power-law particle energy distribution that converges with increasing system size, but the time required to fully form this distribution slowly increases with system size\rm. We suggest that the apparent size dependence of the (fixed-time) index $\alpha$ in our simulations is due to the power laws becoming contaminated by the pileup at $\gamma_{\rm max}$, which develops earlier for smaller systems. We now provide physical motivation for this proposal by considering the particle acceleration process in detail.

   \begin{figure}
  \includegraphics[width=\columnwidth]{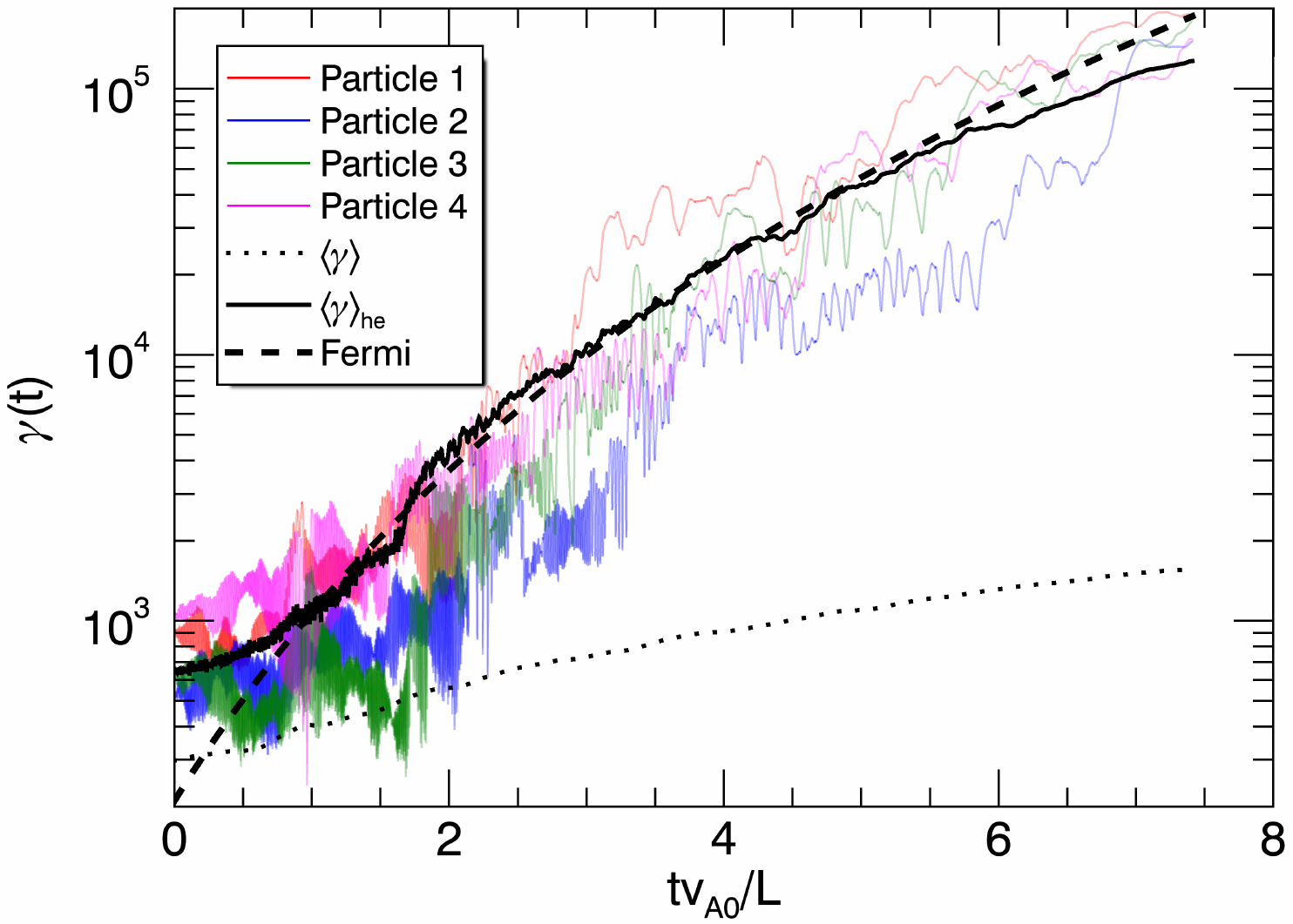}
   \centering
   \caption{\label{fig:tracking} Energy evolution of four particles that attain the highest final energies in the tracked particle sample (solid, colored), along with average of particles that attain $\gamma > 10^5$ (solid, black). The prediction from second-order Fermi acceleration (Eq.~\ref{eq:energy2}; black, dashed) and the overall mean particle energy $\langle\gamma\rangle$ (black, dotted) are also shown for reference.}
 \end{figure}
 
To understand the acceleration process, we tracked a random sample of $8 \times 10^5$ particles in each simulation. In Fig.~\ref{fig:tracking}, we show the energy evolution for the four tracked particles that attain the highest energies at the end of the $1536^3$ simulation. These four particles are the only tracked particles with final energies $\gamma > \gamma_{\rm max}$. At early times, the particle energies exhibit rapid oscillations on a timescale comparable to their Larmor period. The particles occasionally undergo acceleration episodes in which their energy rapidly increases by a factor of 2 or more, but the overall acceleration takes place gradually over several large-scale dynamical times.
 
In the same figure, we show the average energy evolution for all tracked particles with final energies $\gamma > 10^5$ (yielding 44 particles), denoted~$\langle\gamma\rangle_{\rm he}$. After turbulence is fully developed, $\langle\gamma\rangle_{\rm he}$ increases at a slightly sub-exponential rate, until it approaches $\gamma_{\rm max}$. As we now show, this energy evolution is consistent with Fermi acceleration with a slowly evolving acceleration timescale. We suppose that the scattering process causes particle energies to increase as
 \begin{align}
\frac{d\gamma}{dt} \sim \frac{\gamma}{\tau_{\rm acc}} \, , \label{eq:1}
 \end{align}
where $\tau_{\rm acc}(t)$ is the acceleration timescale. For second-order Fermi acceleration, assuming scattering by large-scale fluctuations, the acceleration timescale is
\begin{align}
\tau_{\rm acc} \sim \frac{3}{4} \frac{\lambda_{\rm mfp} c}{u_A^2} \, ,
\end{align}
where $u_A = v_A/(1-v_A^2/c^2)^{1/2} = \sigma^{1/2} c$ is the Alfv\'{e}n four-velocity and $\lambda_{\rm mfp}$ is the scattering mean free path \citep{longair_2011}. For time-independent $\tau_{\rm acc}$, Eq.~\ref{eq:1} leads to an exponential increase in the particle energy. The time required for particles to reach $\gamma_{\rm max}$ from an initial energy of $\gamma_i \sim \gamma_0$ is then $t/\tau_{\rm acc} \sim \log{(\gamma_{\rm max}/\gamma_i)} \sim \log{(L/\rho_{e0})}$. In relativistic plasmas with no energy sink, however, the acceleration timescale evolves with time since the Alfv\'{e}n velocity decreases due to turbulent energy dissipation increasing the relativistic plasma inertia. As discussed in \cite{zhdankin_etal_2018}, for a constant energy injection rate, $\langle\gamma\rangle \sim \gamma_0 (1 + \eta \sigma_0 v_{A0} t/L)$ (where $\eta \approx 1$ is the measured injection efficiency for the given simulations), the Alfv\'{e}n velocity $v_A(t)$ is given by
\begin{align}
\frac{v_A}{c} &= \sqrt{\frac{\sigma}{\sigma + 1}} = \left( 1 + \frac{\langle\gamma\rangle}{\sigma_0 \gamma_0} \right)^{-1/2} \nonumber \\
&\sim \frac{v_{A0}}{c} \left( 1 + \eta \frac{v_{A0}^2}{c^2} \frac{tv_{A0}}{L} \right)^{-1/2} \, .  \label{eq:2}
\end{align}
The energy growth due to second-order Fermi acceleration is then a power law in time (solving Eqs.~\ref{eq:1}-\ref{eq:2}),
\begin{align}
\gamma \sim \gamma_i \left( 1 + \eta \sigma_0 \frac{tv_{A0}}{L} \right)^{4 L c/3 \eta \lambda_{\rm mfp} v_{A0}} \, . \label{eq:energy2}
\end{align}
Note that, using the identity $(1 + x/n)^n \to \exp{x}$ as $n \to \infty$, this equation approaches an exponential in the limit of $\eta \sigma_0 \ll 1$, consistent with time-independent Fermi acceleration. We find that Eq.~\ref{eq:energy2} provides a good fit to $\langle\gamma\rangle_{\rm he}$, as shown in Fig.~\ref{fig:tracking}, if we take $\lambda_{\rm mfp}/L = 1/2$ and $\gamma_i =0.7 \gamma_0$ (giving $\gamma \propto t^{5.1}$ at late times). These results imply that the second-order Fermi process can account for particle acceleration observed in the simulations. First-order acceleration may also contribute; measuring the relative importance of first-order and second-order mechanisms is left for future work.

Inverting Eq.~\ref{eq:energy2} gives the time required for the particle to reach a given energy $\gamma$,
\begin{align}
\frac{t v_{A0}}{L} \sim \frac{1}{\eta \sigma_0} \left[ \left( \frac{\gamma}{\gamma_i} \right)^{3 \eta \lambda_{\rm mfp} v_{A0}/4 L c} - 1 \right] \, . \label{eq:time2}
\end{align}
This equation can be used to estimate the time required for Fermi-accelerated particles to reach the system size limit ($\gamma \sim \gamma_{\rm max}$). Note that Eq.~\ref{eq:time2} approaches a logarithmic function in the limit of $\eta \sigma_0 \ll 1$.

We now relate Fermi acceleration to the late-time evolution of the energy distributions. As previously discussed, the distributions form a power law and then subsequently develop a broad pileup near $\gamma_{\rm max}$. It is natural to focus on the distribution just prior to the pileup formation, when the power law has its maximum extent. After the power law is fully formed, a pair of inflection points appear due to the pileup (which makes the distribution no longer concave down).  We define the inflection time, $t_{\rm inf}$, as the latest time at which the difference between local power-law indices [$\alpha(\gamma) \equiv - \partial\log{f}/\partial\log{\gamma}$] at the two inflection points [local extrema of $\alpha(\gamma)$] is less than $0.1$. For $t > t_{\rm inf}$, the distributions become influenced by the pileup, making a power-law index difficult to define precisely.

\begin{figure}
   \includegraphics[width=\columnwidth]{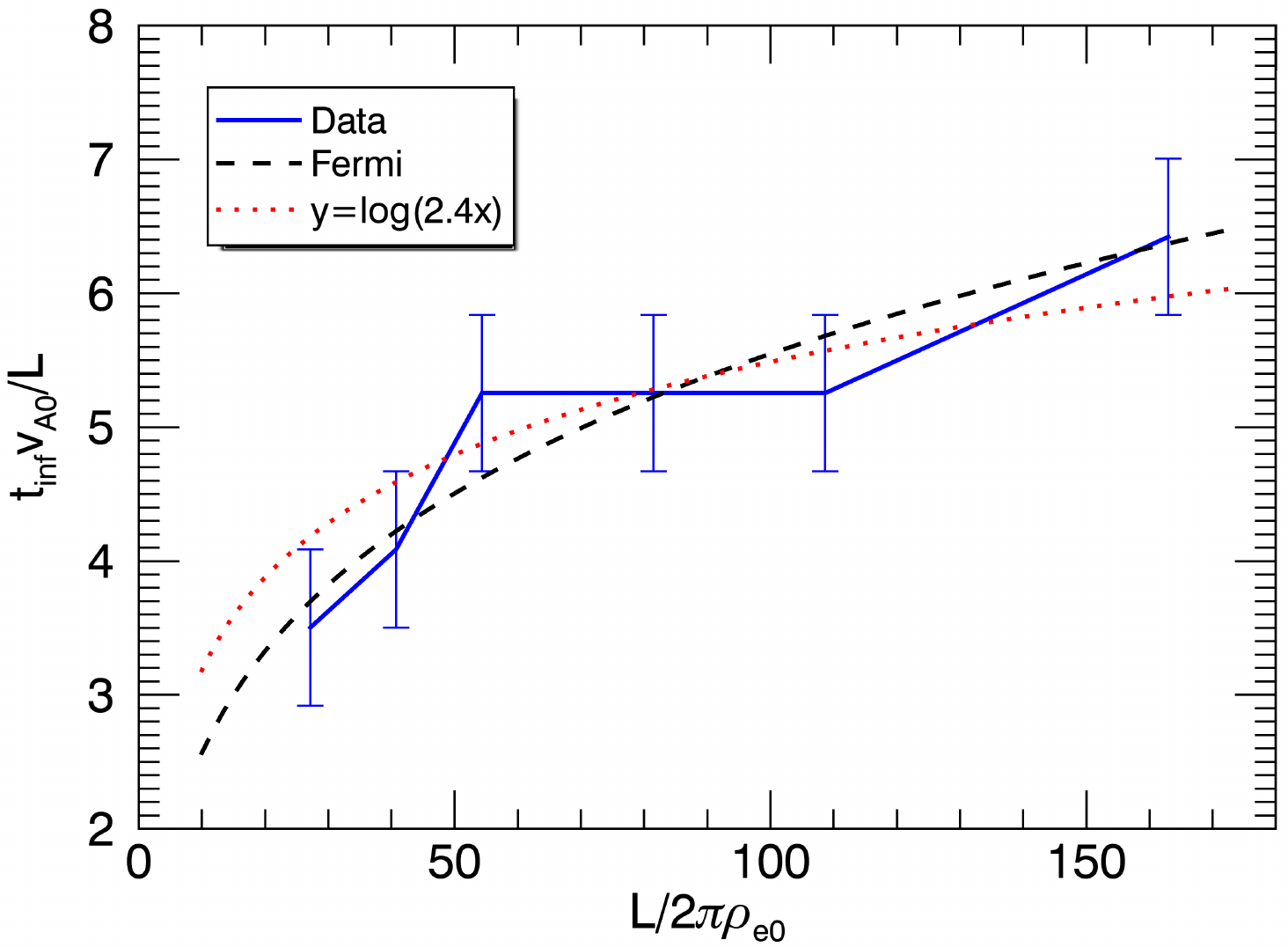}
   \includegraphics[width=\columnwidth]{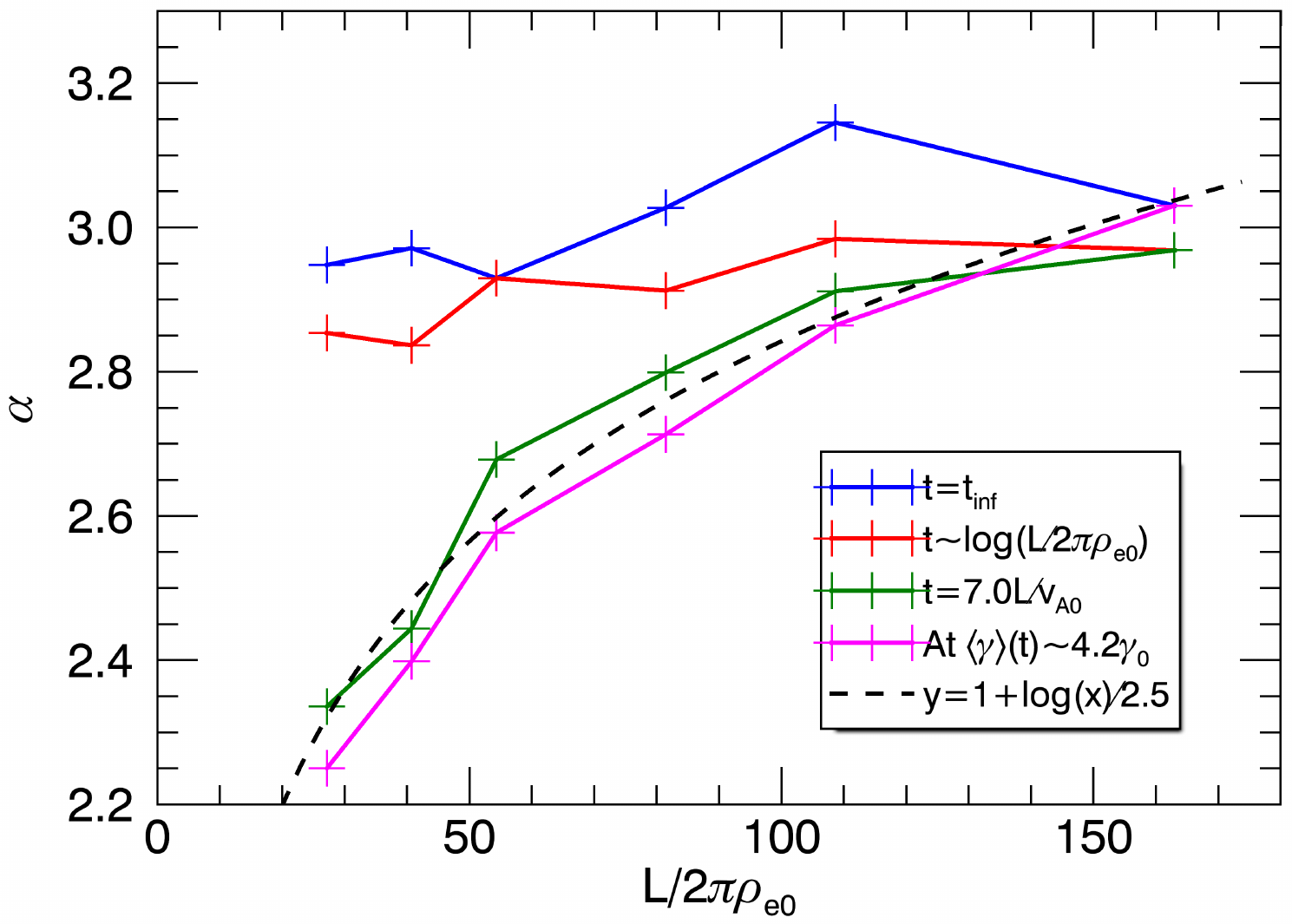}
   \centering
   \caption{\label{fig:alpha_v_size} Top panel: Time taken for the primary inflection point to appear in the energy distribution, $t_{\rm inf}$, versus system size $L/2\pi\rho_{e0}$. The predicted time for particles to be Fermi-accelerated to the system-size limit $\gamma_{\rm max}$ (Eq.~\ref{eq:time2}; black, dashed) and a logarithmic scaling (red, dotted) are also shown. Error bars indicate the time intervals between successive measurements of the distribution. Bottom panel: Power-law index $\alpha$ vs $L/2\pi\rho_{e0}$ measured at various times: at the inflection time $t_{\rm inf}$ (blue), at times scaling logarithmically with size (red), at arbitrary fixed time $t = 7 L/v_{A0}$ (green), and at times with fixed mean particle energy, $\langle\gamma\rangle \sim 4.2 \gamma_0$ (magenta). A logarithmic fit is shown for comparison (black, dashed).}
 \end{figure}

The normalized inflection time $t_{\rm inf} v_{A0}/L$ versus system size $L/2\pi\rho_{e0}$ is shown in the top panel of Fig.~\ref{fig:alpha_v_size}. We find that $t_{\rm inf} v_{A0}/L$ increases with size, consistent with particles requiring a longer number of larger-scale dynamical times to reach $\gamma_{\rm max}$. In fact, $t_{\rm inf}$ is consistent with the second-order Fermi acceleration timescale calculated in Eq.~\ref{eq:time2} [with $L_{\rm mfp}/L = 1/2$ and $\gamma_i = \gamma_0$, giving $t_{\rm inf} \propto (\gamma_{\rm max}/\gamma_0)^{0.2}$], which is also shown the top panel of Fig.~\ref{fig:alpha_v_size}. This scaling is close to logarithmic over the given range of sizes. We note that the time taken for the distribution to reach energies slightly beyond $\gamma_{\rm max}$ exhibits a similar scaling as for $t_{\rm inf}$ (not shown). The system-size dependence of the inflection time, and related pileup, gives a motivation for comparing $f(\gamma)$ at times that increase according to Eq.~\ref{eq:time2}.

In the bottom panel of Fig.~\ref{fig:alpha_v_size}, we show the index $\alpha$ (measured at the logarithmic center of the power law segment) versus system size $L/2\pi\rho_{e0}$ taken at various times: the inflection time $t = t_{\rm inf}$, logarithmic times $t \propto \log{(L/2\pi\rho_{e0})}$, arbitrary fixed time $t = 7 L/v_{A0}$, and at times with fixed mean particle energy $\langle\gamma\rangle \sim 4.2 \gamma_0$ (which is nominally the same as fixed time, but sensitive to statistical variations). We find that measuring the distribution at fixed time or fixed injected energy shows a clear system size dependence, although the dependence weakens with size; in particular, $\alpha$ exhibits an approximately logarithmic dependence on $L/2\pi\rho_{e0}$ [somewhat weaker than suggested in \citep{zhdankin_etal_2017}]. In contrast, the distribution taken at the inflection time or at logarthmic times shows no systematic variation for $L/2\pi\rho_{e0} \gtrsim 80$. The sum up, distributions attain the same power-law index, independent of system size, just prior to pileup formation.
 
We conclude with a comment about the statistical significance of our results. In our simulations, the amount of energy injected into the plasma by the external driving fluctuates randomly in time, since driven mode phases are evolved randomly. While the mean energy injection rate approaches a universal value for sufficiently long simulations, the distributions presented in this paper were measured after a limited duration ($\lesssim 7 L/v_{A0}$) and thus the amount of injected energy at that point can vary significantly between different runs (by up to $\sim 30\%$). In principle, a larger injection of energy may supply a harder nonthermal population, bringing up an important question: do the measured nonthermal distributions exhibit significant statistical variability (from run to run) due to random driving? To build confidence that our largest simulations are not statistical outliers, we analyzed ensembles of sixteen $384^3$ ($L/2\pi\rho_{e0} = 40.7$) and eight $768^3$ ($L/2\pi\rho_{e0}= 81.5$) simulations. For the $384^3$ ensemble, we obtain an average fixed-time index of $\langle\alpha\rangle \approx 2.53$ and rms spread of $\delta\alpha_{\rm rms} \approx 0.06$ (at $t = 7 L/v_{A0}$), with $\alpha$ weakly correlated with injected energy. For the $768^3$ ensemble, we find $\langle\alpha\rangle \approx 2.79$ and $\delta\alpha_{\rm rms} \approx 0.07$, indicating that the statistical spread is similar at both sizes and is less than the difference due to size. When measured at logarithmic times, we find $\langle\alpha\rangle \approx 2.86$, $\delta\alpha_{\rm rms} \approx 0.09$ for the $384^3$ ensemble (at $t v_{A0}/L \approx 4.2$) and $\langle\alpha\rangle \approx 2.94$, $\delta\alpha_{\rm rms} \approx 0.06$ for the $768^3$ ensemble (at $t v_{A0}/L \approx 5.2$), in agreement with Fig.~\ref{fig:alpha_v_size}. In addition to this, we find that the variations are modest in each pair of equal-size simulations from our main system-size scan. Hence, we believe that trends measured in our simulations are robust.

\section{Conclusions}

Based on the simulations presented in this paper, we are prepared to declare the system-size independence of power-law indices of particle energy distributions produced by driven, large-scale turbulence in relativistic collisionless plasmas. We empirically observe that convergence of the power-law indices occurs at times that depend on system size (approximately logarithmically or as a weak power law). We propose a physical interpretation for this dependence of time on system size: as the size is increased, there is an increasing number of scatterings required for particles to acquire the highest accessible energies via Fermi acceleration. The energy distribution becomes fully developed only once particles reach the energy limit $\gamma_{\rm max}$ due to finite system size, and subsequently the distributions experience a high-energy pileup that may complicate measurements of the power law. This pile-up forms due to constant energy injection into a closed system, which is an unrealistic scenario; in general, particle energies may be limited by other physics, such as radiative cooling or open boundaries.

This work demonstrates that accurately measuring the converged power-law index is feasible with modestly large PIC simulations of turbulence (with $L/2\pi\rho_{e} \gtrsim 80$). We find a converged index of $\alpha \approx 3.0$ at magnetization $\sigma = 3/8$ and turbulence amplitude $\delta B_{\rm rms} \sim B_0$; as discussed in \cite{zhdankin_etal_2017}, turbulent particle acceleration becomes more efficient with increasing $\sigma$, leading to smaller values of $\alpha$ and faster formation of the nonthermal population (as implied by Eq.~\ref{eq:time2}). A similar converged value of the index, for comparable plasma parameters, is measured in PIC simulations of relativistic magnetic reconnection \citep[e.g.,][]{werner_uzdensky_2017}; slightly steeper indices are measured for relaxation of magnetostatic equlibria \citep{nalewajko_etal_2016}. Likewise, our measured index is very close to the index of $\approx 3.2$ inferred from the continuum synchrotron spectrum in the Crab nebula, where a magnetization below unity is expected \citep{meyer_etal_2010}.

%% If you wish to include an acknowledgments section in your paper,
%% separate it off from the body of the text using the \acknowledgments
%% command.
\acknowledgments

The authors acknowledge support from NSF grant AST-1411879 and NASA ATP grants NNX16AB28G and NNX17AK57G. An award of computer time was provided by the Innovative and Novel Computational Impact on Theory and Experiment (INCITE) program. This research used resources of the Argonne Leadership Computing Facility, which is a DOE Office of Science User Facility supported under Contract DE-AC02-06CH11357. This work also used the Extreme Science and Engineering Discovery Environment (XSEDE), which is supported by National Science Foundation grant number ACI-1548562. This work used the XSEDE supercomputer Stampede2 at the Texas Advanced Computer Center (TACC) through allocation TG-PHY160032 \citep{xsede}.

\end{document}